\documentclass{PoS}
\usepackage{epsfig}
\usepackage{cite}

\newcommand{\kspll}{K_{S}\rightarrow{\pi^0}l^+l^-}
\newcommand{\klpll}{K_{L}\rightarrow{\pi^0}l^+l^-}
\newcommand{\ksspll}{K_{1}\rightarrow{\pi^0}l^+l^-}

\newcommand{\klpee}{K_{L}\rightarrow{\pi^0}e^+e^-}

\newcommand{\klpmm}{K_{L}\rightarrow{\pi^0}\mu^+\mu^-}
\newcommand{\klpgg}{K_{L}\rightarrow{\pi^0}\gamma\gamma}

\newcommand{\bea}{\begin{eqnarray}}
\newcommand{\eea}{\end{eqnarray}}

\title{Rare Kaon Decays $K\to \pi \nu \bar \nu$ and $K_L\to \pi^0
\ell^+\ell^-$}

\ShortTitle{Rare Kaon Decays $K\to \pi \nu \bar \nu$ and $K_L\to \pi^0
\ell^+\ell^-$}

\author{Federico Mescia\\
        INFN, Laboratori Nazionali di Frascati, Via E.
          Fermi 40, I-00044 Frascati, Italy\\
        E-mail: \email{mescia@lnf.infn.it}}

\abstract{Over the next years, the Flavour Physics community will be
looking for inconsistencies of the Standard Model (SM) by exploiting new and precise
measurements. In these ``indirect new physics'' searches,  the key
strategy is to concentrate on observables, which are theoretically clean
and preferably suppressed in the SM.  In this respect, the four golden
modes  $K^+\to \pi^+ \nu \bar \nu$, $K_L\to \pi^0\nu \bar \nu$ and
$K_L\to \pi^0 \ell^+\ell^-$ are very promising.The pollution
from hadronic uncertainties in these decays is at present under control with 
good  accuracy, thanks  to the interplay between theory information
(Chiral Perturbation Theory and  OPE) and experimental inputs (KTeV,
NA48). Their measurements might thus give rise to some
unexpected scenario. Here, we briefly review the present situation for
these four exclusive rare decays.}

\FullConference{International Europhysics Conference on High Energy Physics\\
                 July 21st - 27th 2005\\
                 Lisboa, Portugal}

\begin{document}

\section{Introduction}
\vspace*{-0.5cm}
The present theoretical and experimental scenario for the four
 modes discussed here is sketched in the table below:
\begin{center}
\vspace*{-0.5cm}
\begin{figure}[h]
\epsfig{file=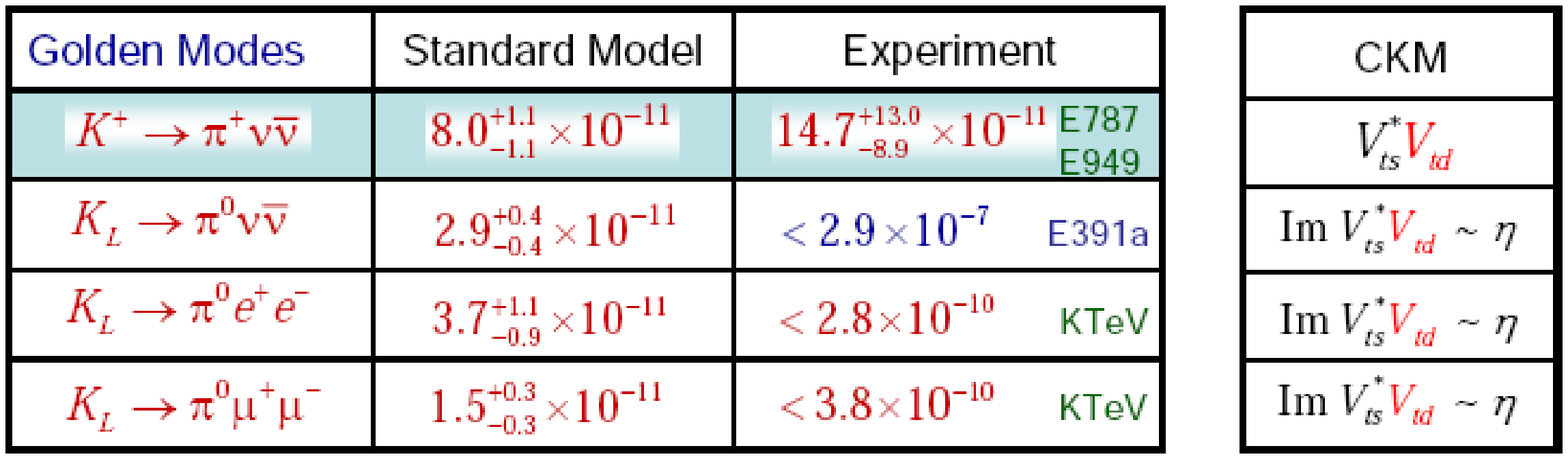,width=1.0\textwidth}
\vspace*{-1.6cm}
\end{figure}
\end{center}
Even though difficulties to measure these decays (with missing energy and
branching ratios below $10^{-10}$) are very high, three events for $K^+\to \pi^+ \nu \bar
\nu$ have been already observed at BNL by the E787-E949 experiments~\cite{E787E949}. The
corresponding measure turns out to be twice higher than the SM, but  
compatible within the large error. For the
future,  experimental progress is connected to the recent NA48 
proposal. Their project is to collect about 80 $K^+\to \pi^+ \nu \bar
\nu$ events 
 in two years starting from 2010. For the other modes, the experimental
 situation is less settled down. For the $K_L\to \pi^0 \nu \bar\nu$ decay, 
 the E391a Collaboration has
improved, this summer~\cite{Kaon5:e391},  the upper limit fixed
earlier by KTeV.
They have analyzed  $10\%$ of their first run and the third one is
being finished. At the end, they plan a $10^{-10}$ sensitivity close to
the SM limit and to possible surprises. Moreover, E391a is intended to be
a pilot experiment at JPARC for a more ambitious project of ${\cal
O}(100)$ $K_L\to \pi^0 \nu \bar\nu$ events. For the two other modes, the
best experimental knowledge is from KTeV but no new
experiments are planned yet. The advantage of these decays is that
all the decay products can be detected and this should not be ignored. 

From the theoretical point of view, these decays arise from FCNC
one-loop transitions. For this reason, they provide an independent
scrutiny of  $V_{td}$ from  unexplored sectors:   $\Delta S=1$ at  
one-loop level. Indeed, the flavor dynamics~\cite{utfit}
has been mostly investigated  on the $\Delta
B=1\,,2$ and $\Delta S=2$ transitions.
Due their theoretical cleanness, they could also shed light on the
nature of new
physics~\cite{Hurth:2005wg,GinoCern,D'Ambrosio:2001zh,Bobeth:2005ck},
when new particles will be directly observed at LHC.

In what follows, we will review the theoretical status of these
decays, with emphasis on recent progress to reduce 
the hadronic uncertainties. 
\vspace*{-0.4cm}

\section{Hadronic uncertainties: Long Distance effects and power
Corrections}
\vspace*{-0.5cm}

The short-distance (SD) contributions  to these decays are described 
at leading order~\cite{Buras1}  by
\vspace*{-0.3cm}
\bea
H_{eff}=\frac{G_{F}}{\sqrt{2}}\sum_{q=u,c,t}\lambda_{q}\left(c^q_{\nu}(\bar{s}d)_{V-A}(\nu\bar{\nu})_{V-A}+
 c^q_{V}(\bar{s}d)_{V-A}(\ell^{+}\ell^{-})_{V}
+ c^q_{A}(\bar{s}d)_{V-A}(\ell^{+}\ell^{-})_{A}\right)  
\label{eq:heff}
\eea
with $\lambda_{q}=V_{qs}^{\ast}V_{qd}$. 
$c^q_{\nu}$ and $c^q_{A}$ encode W-boxes and Z-penguins, 
whereas $c^q_{V}$ Z- and $\gamma-$ penguins.

Due to the power-like GIM enhancement in Z- and W-mediated processes, 
top loops represent  the leading contributions to eq.~(\ref{eq:heff})  
and are computable in perturbation theory with  high precision. 
The hadronic uncertainty then due to $\langle\pi\vert(\bar s d)_{V-A}\vert K\rangle$ 
in eq.~(\ref{eq:heff})  is accurately  determined  by 
the $K_{\ell3}$ rates (modulo isospin
corrections). According thus to the relative strength of the top contribution, these decays
are more or less theoretically clean. Being for example $K_L\to \pi^0\nu \bar
\nu$ a CP-violating (CPV) process,  the GIM mechanism along with the CKM
hierarchy makes CPV effects from  charm and up completely 
negligible~\cite{buca}, $\le 1\%$,
and the corresponding
decay turns out to be the cleanest of these four
modes.  Unlike $K_L\to \pi^0\nu \bar \nu$ charm and up loops can
however contribute to the others through  CPV-effects mediated by
$\gamma$~\footnote{For the $\gamma$-penguins, the GIM mechanism
is only logarithmic, so that up and charm can be sizable, 
namely $\log(m_q/M_W)\gg\log(m_t/M_W)$.}  and/or CP-conserving
(CPC) effects mediated 
by $\gamma$ and Z. In this case, for both 
$K^+\to \pi^+ \nu \bar \nu$ and $K_L\to \pi^0 \ell^+\ell^-$ charm and up
can induce hadronic uncertainties in terms of ${\cal O}(G_F)$
electro-weak corrections to the non-leptonic $\Delta S=1$ 
operators. At present, all these contributions 
are small and/or under good theoretical control. Moreover, 
 their present knowledge  could be also improved
by means of lattice QCD~\cite{turco}.

In the case of CP-conserving transition $K^+\to\pi^+\nu\bar\nu$ for
example,  the top 
enhancement with respect to the charm is partially compensated  by the  CKM
coefficient ($|V^{\ast}_{cs}V_{cd}| \approx 10^3 \times  |V^{\ast}_{ts}V_{td}|$)
and the charm then amounts to about $35\%$ of the total magnitude.  
Due to the low value of the charm mass, close to the non-perturbative 
regime, either NNLO QCD corrections for the coefficient $y^c_\nu$
in~(\ref{eq:heff}) or subleading terms $\propto 1/m_c^2$ not described by the
 Hamiltonian in eq.~(\ref{eq:heff}) cannot be completely neglected. 
The NNLO calculation has been recently performed in~\cite{Buras:2005gr}
and $y^c_\nu$
is now known with a relative precision of about $9\%$. This 
translates into an error  of about $5\%$ in the SM estimate of
$Br(K^+\to\pi^+\nu\bar\nu)$.  
For the subleading terms, we have  both dimension-eight
four-fermion operators generated at the charm  scale, and genuine
$\Delta S=1$ long-distance contributions which   can be described within
the framework of Chiral Perturbation  theory (ChPT). In~\cite{Isidori:2005xm},
we have shown that  a consistent treatment of the latter  contributions,
which turn out to be the dominant effect, requires the introduction of
new chiral operators already  at $O(G_F^2 p^2)$. 
Using this new chiral Lagrangian, an approximate matching between short- and
long-distance components has been performed and from the numerical point of
view,  these   corrections enhance the SM
prediction of $Br(K^+\to\pi^+\nu\bar\nu)$ by about $6\%$. A residual 

Finally, in the case of $\klpll$,  the SD Hamiltonian in~(\ref{eq:heff})
takes into account only the direct CPV effects  
from top ($\sim i \lambda^5 m_t^2/m^2_W$) and 
charm ($\sim i \lambda^5 \log(m_c^2/m^2_t)$) loops. However, because of
the presence of
the up contributions  mediated by photon penguins,  the
$\klpll$ modes receive also two long-distance contributions:  indirect
CP-violation (ICPV) and CPC contributions. In addition, DCPV and ICPV components
interfere with each other and the sigh of this interference contribution
(INTF)
has to be fixed as well.
More specifically,   through
the $K^0-\overline{K^0}$ mixing,  the CPC decay $\ksspll$ gives rise
to   a ICPV contribution to the $\klpll$, computable in
ChPT~\cite{ambrosio98} using the  recent measurements of the
$\kspll$ modes~\cite{NA48as}.  A constructive interference is favored now by
two works~\cite{bdi,friot}. 

The CPC contribution instead proceeds through two virtual photons and
has been  recently determined~\cite{bdi,isu} by a precise study of the decay
$\klpgg$~\cite{klpgg}. 
Finally, the pattern for the $BR$'s in the SM ($\rm{Im}\lambda_t\sim 1.36
\,10^{-4}$) is:
\vspace*{-0.4cm}
\bea
BR(\klpee)_{SM} &\approx& 
(23_{ICPV} + 10_{INTF} +
4_{DCPV})\times 10^{-12}\,, \\
BR(\klpmm)_{SM} &\approx&
(5.4_{ICPV} + 2.5_{INTF} + 2_{DCPV}
+ 5_{CPC})\times 10^{-12}\,,
\eea
where the irreducible theoretical error on the various contributions is around
$10\%$. The INTF term is prortional to $\rm{Im}\lambda_t$, whereas the
DCPV to $\rm{Im}\lambda_t$ squared.
This means that despite of dominance of the long-distance indirect
CP-violating contribution  we are still able to uncover with relatively
good accuracy  the short-distance part related to $\lambda_t$ (and possibly
sensitive to new physics).

\vspace*{-0.2cm}
\section{Summary and Prospects}
\vspace*{-0.4cm}
In conclusion, estimating all the hadronic uncertainties (LD effects) with good
accuracy, SD contributions and thus potential new physics effects can be
clearly unveiled. In perspective, these decays  
are as promising as $\sin 2\beta$ from $B\to J/\psi
K_S$ in the B-factory era.
\vspace*{-0.4cm}

\section{Acknowledgments} 
\vspace*{-0.4cm}
We thank G.~Isidori, P.~Paradisi, C.~Smith and S.~Trine  for discussions on
the subject  of this talk.  This work has been supported by IHP-RTN,  EC
contract No. HPRN-CT-2002-00311 (EURIDICE).


\begin{thebibliography}{99}

\bibitem {E787E949}S.~Adler \textit{et al. }[E787], Phys. Rev. Lett.
\textbf{88} (2002) 041803; V.~V.~Anisimovsky \textit{et al. }[E949], Phys.
Rev. Lett. \textbf{93} (2004) 031801.

\bibitem{Kaon5:e391}
K.~Sakashita, talk given at the Kaon 2005 International Workshop,
http://www-ps.kek.jp/e391/kaon2005\_ e391a\_kpi0nn.pdf. 

\bibitem{utfit}
M.~Bona {\it et al.}  [UTfit Collaboration],
JHEP {\bf 0507}, 028 (2005)
[arXiv:hep-ph/0501199].\\
  J.~Charles {\it et al.}  [CKMfitter Group],
  Eur.\ Phys.\ J.\ C {\bf 41}, 1 (2005)
  [arXiv:hep-ph/0406184].

\bibitem{Hurth:2005wg}
  T.~Hurth,
  arXiv:hep-ph/0511280.

\bibitem{GinoCern}
Talk of G.~Isidori at workshop, ``Flavour physics in the era of LHC '',
November 6th-12th, Cern, Geneva.


\bibitem{D'Ambrosio:2001zh}
G.~D'Ambrosio and G.~Isidori,
Phys.\ Lett.\ B {\bf 530}, 108 (2002)
[arXiv:hep-ph/0112135].

\bibitem{Bobeth:2005ck}
C.~Bobeth, M.~Bona, A.~J.~Buras, T.~Ewerth, M.~Pierini, L.~Silvestrini and A.~Weiler,
Nucl.\ Phys.\ B {\bf 726}, 252 (2005)
[arXiv:hep-ph/0505110].


\bibitem{Buras1}A.J.~Buras, M.E.~Lautenbacher, M.~Misiak and M.~M\"{u}nz, Nucl.
Phys. \textbf{B423} (1994) 349; G. Buchalla, A.J. Buras and M.E. Lautenbacher,
Rev. Mod. Phys. \textbf{68} (1996) 1125.

\bibitem{buca}
G.~Buchalla and G.~Isidori,
Phys.\ Lett.\ B {\bf 440}, 170 (1998)
[arXiv:hep-ph/9806501].

\bibitem{turco}
G.~Isidori, G.~Martinelli and P.~Turchetti,
arXiv:hep-lat/0506026.


\bibitem{ambrosio98}
G.~D'Ambrosio, G.~Ecker, G.~Isidori and J.~Portoles,
JHEP {\bf 9808}, 004 (1998)
[arXiv:hep-ph/9808289].

\bibitem{Buras:2005gr}
A.~J.~Buras, M.~Gorbahn, U.~Haisch and U.~Nierste,
arXiv:hep-ph/0508165;
 U.~Haisch,
arXiv:hep-ph/0512007.

\bibitem{Isidori:2005xm}
G.~Isidori, F.~Mescia and C.~Smith,
Nucl.\ Phys.\ B {\bf 718}, 319 (2005)
[arXiv:hep-ph/0503107]; C.~Smith, arXiv:hep-ph/0505163.

\bibitem {NA48as}J.R.~Batley et al. [NA48], Phys. Lett. \textbf{B576} (2003)
43; Phys. Lett. \textbf{B599} (2004) 197.

\bibitem{bdi} G.~Buchalla, G.~D'Ambrosio and G.~Isidori,
Nucl.\ Phys.\ B {\bf 672}, 387 (2003)
[arXiv:hep-ph/0308008].

\bibitem{friot}
S.~Friot, D.~Greynat and E.~De Rafael,
Phys.\ Lett.\ B {\bf 595}, 301 (2004)

\bibitem{isu}
G.~Isidori, C.~Smith and R.~Unterdorfer,
Eur.\ Phys.\ J.\ C {\bf 36}, 57 (2004)
[arXiv:hep-ph/0404127]; C.~Smith, arXiv:hep-ph/0407361.
.


\bibitem {klpgg}
 A.~Alavi-Harati et al. [KTeV], Phys. Rev. Lett.
\textbf{83} (1999) 917;\\
A.~Lai et al. [NA48], Phys. Lett. \textbf{B536} (2002) 229.


\end{thebibliography}
\end{document}